\documentclass[useAMS]{mn2e}
\usepackage{graphics}
\usepackage{rotate}
\usepackage{times}
\newif\ifAMStwofonts
\AMStwofontstrue
\input{psfig.sty}

\def\gsim{\mathrel{\hbox{\rlap{\hbox{\lower4pt\hbox{$\sim$}}}\hbox{$>$}}}}
\def\lsim{\mathrel{\hbox{\rlap{\hbox{\lower4pt\hbox{$\sim$}}}\hbox{$<$}}}}

\begin{document}

\title[XTE~J1650--500: evidence for BH spin and light bending effects?]  
{The relativistic Fe emission line in XTE~J1650--500 with
  {\it{BeppoSAX}}: evidence for black hole spin and light bending effects?}  
  \author[G. Miniutti, A.C. Fabian, J.M. Miller ] {G.
  Miniutti$^1$\thanks{E-mail: miniutti@ast.cam.ac.uk}, A.C.
  Fabian$^{1}$, J.M. Miller$^{2,3}$ \\ $^1$ Institute of Astronomy,
  University of Cambridge, Madingley Road, Cambridge CB3 0HA \\ $^2$
  Harvard--Smithsonian Center for Astrophysics, 60 Garden Street,
  Cambridge MA 02138, USA  \\ $^3$ NSF Astronomy and Astrophysics
  Postdoctoral fellow}


\pubyear{2003}

\maketitle

\label{firstpage}

\begin{abstract}
{ We report spectral results from three {\it{BeppoSAX}}
observations of the black hole candidate XTE~J1650--500 during its
2001/2002 outburst. We find strong evidence for the presence of a
broad and strongly relativistic Fe emission line. The line profile
indicates an accretion disc extending down to two gravitational radii
 suggesting the presence of a rapidly rotating central Kerr
black hole. Thanks to the broadband spectral coverage of
{\it{BeppoSAX}}, we could analyse the 1.5--200~keV spectra of the
three observations and report the presence of a strong reflection
component from the accretion disc, which is totally consistent with the
observed broad Fe emission line. The shape of the reflection component
appears to be affected by the same special and general relativistic
effects that produce the broad Fe line. We study the variation of the
different spectral components from the first to the third observation
and we find that they are well reproduced by a recently
proposed light bending model.}
\end{abstract}

\begin{keywords} black hole physics --- X-rays: individual 
   (XTE~J1650--500) --- X-rays: stars 
\end{keywords}

\section{Introduction}

Black Hole Candidates (BHC) often exhibit transitions between
different X--ray states, defined by their different spectral and
temporal behaviour \cite{tl95,vdk95,mk97,mcrem03}. Transitions between
different X--ray states are believed to be mainly due to variations in
the mass accretion rate. However, recent observations (see e.g. Homan
et al. 2001) strongly suggest that a single parameter is not
sufficient to explain the nature of the different states, and at least
a second one must be considered, whose nature is still unclear. Black
hole X--ray binaries appear to evolve through a continuous range of
states whose general properties strongly depend mainly on the relative
contribution and interplay between two main spectral components, i.e.
a soft thermal disc component and a hard non--thermal power law
component. In some cases the presence of a reflection component and/or
of emission lines and edges has been observed suggesting reprocessing
of the hard X--rays by cold material (generally identified with the
accretion disc).

Here we report spectral results from three {\it{BeppoSAX}}
observations of the BHC XTE~J1650--500 during its 2001/2002 outburst.
The X--ray transient XTE~J1650--500 was first detected in outburst
with the {\it{Rossi X-Ray Timing Explorer}} ({\small{\it{RXTE}}}) on
2001 September 5 \cite{remi01}. Optical and radio counterparts were
identified by Castro--Tirado et al. (2001) and Groot et al. (2001),
respectively.  Subsequent observations revealed X--ray variability, a
hard spectrum and quasi--periodic oscillations (QPOs) at a few Hz
making the source a BHC , although the mass of the central object is
still not well constrained by the data \cite{mss01,rs01,wml01}. High
frequency variability at about 250~Hz (and also likely at 250$\times$
2/3 Hz) has also been detected \cite{homan03}. If the 250~Hz
oscillation is interpreted as the orbital frequency at the innermost
stable circular orbit, one can obtain a mass estimate of 8.2 M$_\odot$
for a non--rotating black hole. If a rotating black hole is
considered, the black hole mass could well be much higher.

{\it{XMM--Newton}} observations close to the peak of the outburst in
the very high state revealed the presence of a broad and asymmetric Fe
emission line suggesting that the black hole in XTE~J1650--500 is 
rapidly rotating \cite{miller02a}. Fe line diagnostics
is potentially very important for constraining the accretion flow
geometry, black hole spin and the nature and location of the primary
source illuminating the accretion disc. Such diagnostics has already
been used in both BHC (see for example Martocchia et al. 2002;
Miller et al. 2003, 2002a,b,c) and active galaxies 
(see e.g. Tanaka et al. 1995; Wilms et al. 2001; 
Fabian et al. 2002 for the most remarkable case of MCG--6-30-15).

The {\it{BeppoSAX}} observations of the 2001/2002 XTE~J1650--500
outburst were performed just before and after the {\it{XMM--Newton}}
observation that revealed the presence of a relativistic Fe emission
line. Our main purpose is to explore the Fe line energy band to
confirm (or not) the {\it{XMM--Newton}} detection and to obtain
improved constraints from the broadband spectra; thanks to
the broadband coverage of the {\it{BeppoSAX}} instruments, we present a
self--consistent analysis of the 1.5--200~keV spectrum XTE~J1650--500
during its 2001/2002 outburst.

\section{Observations}

XTE~J1650--500 was observed 3 times by the Italian--Dutch satellite
{\it{BeppoSAX}} \cite{betal97a} during 2001. Here we report on the
three observations made on September 11 (obs.~1), September 21
(obs.~2) and October 3 (obs.~3).  For comparison, we note that the
{\it{XMM--Newton}} observation by Miller et al. 2002 that revealed the
presence of a broad, relativistic Fe line was performed on September
13. In this paper, data from the imaging instruments (LECS Parmar et
al. 1997 and MECS Boella et al. 1997b) and from the collimated PDS
\cite{fetal97} instrument are reported. The data files have been
obtained from the {\it{BeppoSAX}} public archive.  Spectra for the
LECS and MECS instruments are obtained within a circular region
centred on the source with radius of 8 arcmin. The background was
extracted from event files of source--free regions of the same
size. Spectra were grouped such that each bin contains at least 20 counts.
Exposures times for LECS were of 20~ks, 14~ks and 7~ks (obs.
1,2 and 3 respectively), while the MECS exposures were of 47~ks,
64~ks and 28~ks. The PDS exposure times were of 21~ks, 31~ks
and 12~ks. The {\it{BeppoSAX}} data were fitted by using the
{\small{XSPEC}} 11.2 package \cite{arnaud}. In the following, all the quoted
uncertainties on the parameters correspond to 90 per cent confidence
intervals for one interesting parameter ($\Delta \chi^2 =2.71$).

\section{Spectral analysis}

We first analysed the MECS data in the 2.5--10~keV energy band in
order to look for the presence of a broad Fe emission line as observed
in the September 13 {\it{XMM--Newton}} observation by Miller et al.
(2002a). Some feature is present in the MECS data below 2.5~keV,
independently of the spectral model that is used to analyse the data.
We thus decided to ignore the low energy data below 2.5~keV and focus
on the iron line energy band. To better constrain the continuum and to
assess the relevance of reflection components in the spectra, we then
added the PDS data in the range 13--200~keV. Some data from the LECS
instruments (~1.5--3.0~keV~) have also been considered and we report
our analysis of the broadband 1.5--200~keV data
(~LECS~+~MECS~+~PDS~).  

\subsection{The 2.5--10~keV spectrum from the MECS instrument}

\begin{figure*}
\begin{center}
\centerline{\mbox{ 
\psfig{figure=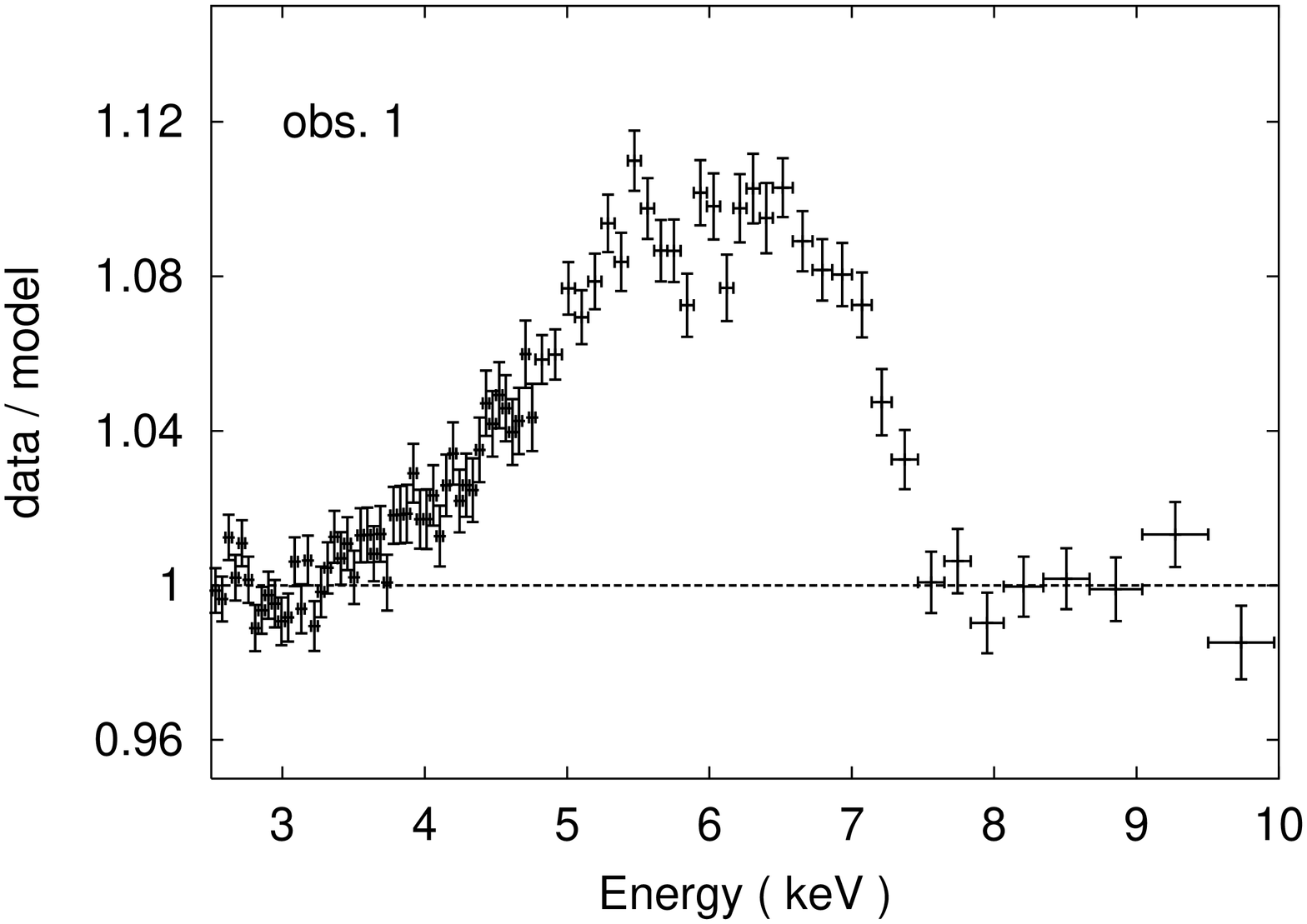,width=5.5cm,height=4.0cm,angle=-90}
\hspace{0.2cm}
\psfig{figure=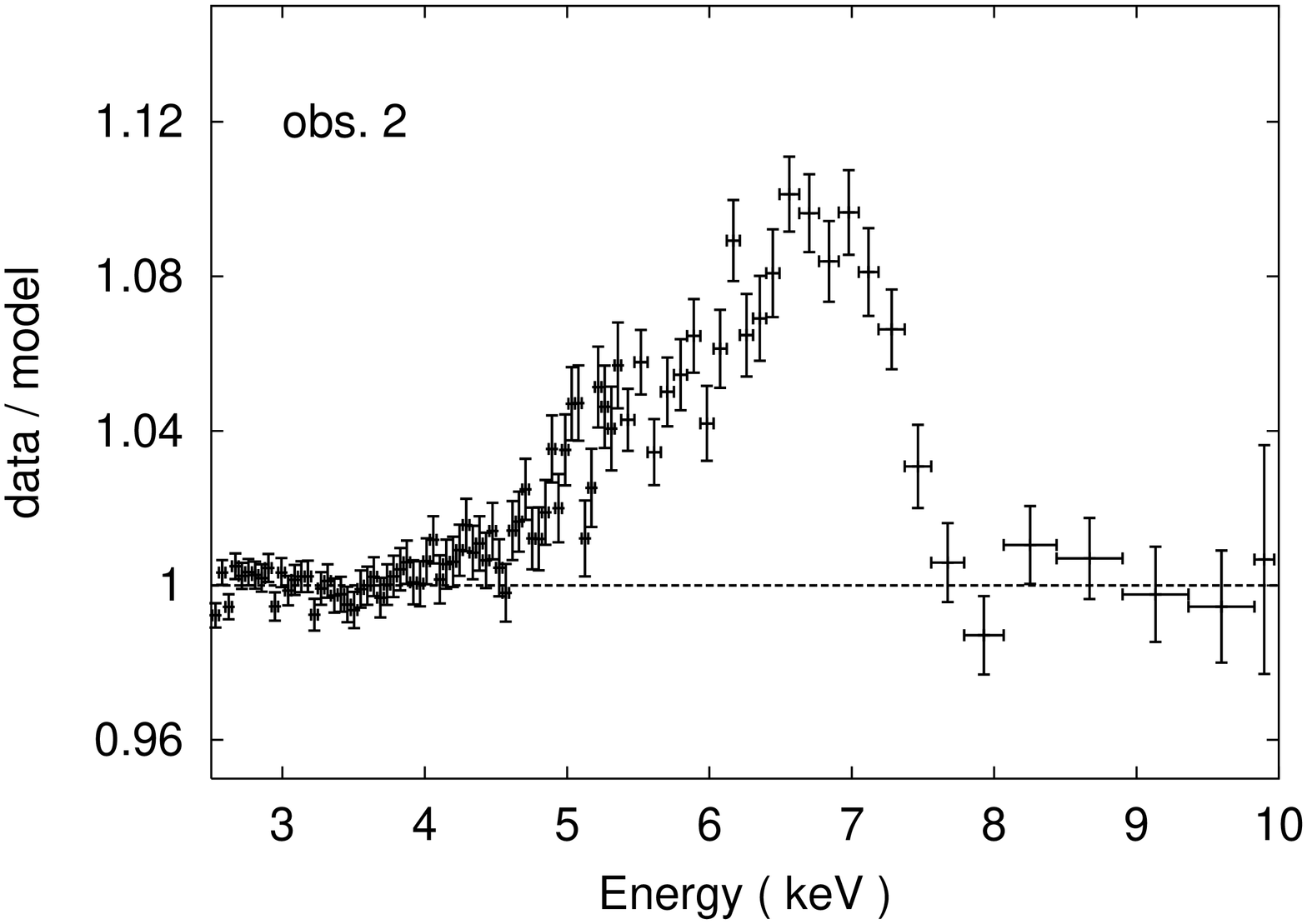,width=5.5cm,height=4.0cm,angle=-90}
\hspace{0.2cm}
\psfig{figure=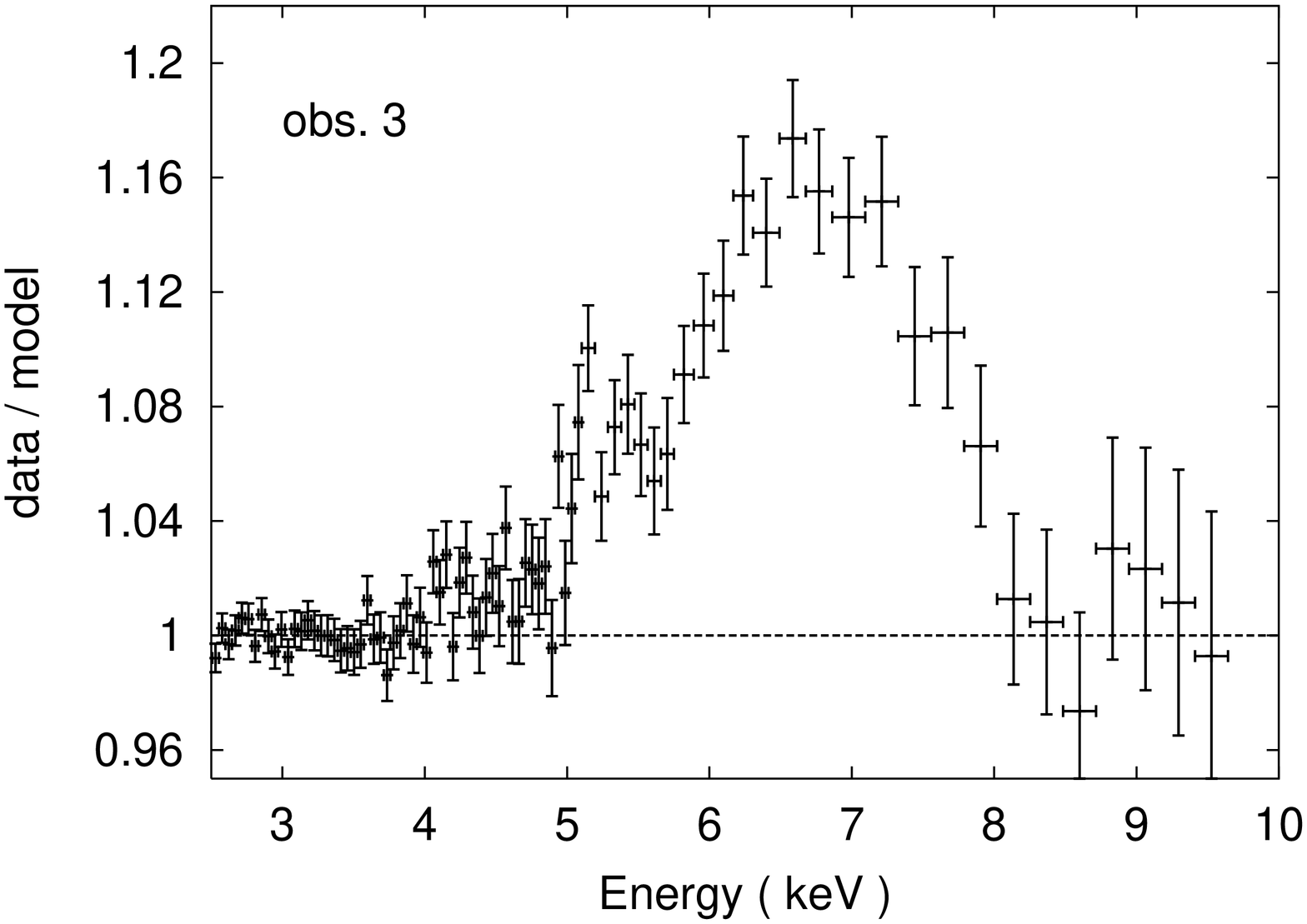,width=5.5cm,height=4.0cm,angle=-90} 
}}
\vspace{0.2cm}
\caption{2.5--10~keV data/model ratio
  obtained with a multicolour disc black body component, a power law
  and a smeared edge modified by photoelectric absorption for the
  three {\it{BeppoSAX}} observations. The 4--8~keV band, where
  emission line features are expected, has been ignored in fitting the
  model. Ratios have been rebinned for visual clarity.  }
\end{center}
\end{figure*}
We first analysed the MECS data by considering the ``standard'' model for
galactic black hole candidates X--ray sources, i.e. a multicolour
accretion disc blackbody (MCD) from Mitsuda et al. (1984) and a
power law, modified by absorption in the interstellar medium (the
{\small{PHABS}} model in {\small{XSPEC}}). The fit is very poor for
the first two observations ($\chi^2 = 958$ and $\chi^2 = 422$,
respectively) and better for obs. 3 ($\chi^2 = 200$) for 156 degrees
of freedom. The main reason for the low quality of the fits is the
presence of large residuals in the region of the Fe complex, above
3.5--4~keV. The shape of the residuals strongly suggest the presence
of a broad and asymmetric Fe emission line in the three
observations. 

We then describe the 2.5--10~keV MECS data by adding the
{\small{LAOR}} model to account for relativistic iron line emission
from an accretion disc around a Kerr black hole. In this model, the
emissivity profile on the accretion disc is described by a power law
of the form $\epsilon (r) = r^{-\beta}$, where the emissivity index
$\beta$ is a free parameter of the model. After some testing, the
inclination was fixed to 45 degrees (as in Miller et al. 2002) as the
data indicate clear preference for intermediate inclinations. The
outer disc radius was fixed to its maximum possible value of
$400~r_g$, where $r_g = G~M/c^2$ and $M$ is the black hole mass. The
inner disc radius, the emissivity index, the rest energy of the
emission line, and the model normalisation were let free to vary
during the fit. A smeared edge ({\small{SMEDGE}}) was also added to
the model. This phenomenological model for the absorption edge is here
considered for comparison with previous works only. A more
physically--motivated reflection component will be considered in the
next Section where the 1.5--200~keV spectra are discussed.

The addition of the {\small{SMEDGE}} and {\small{LAOR}} components
significantly improved the fit ($\chi^2 = 158$, $\chi^2 = 197$, and
$\chi^2 = 149$ for obs. 1, 2, and 3 respectively, for 149 degrees of
freedom) with respect to our initial model
(~{\small{DISKBB~+~POWERLAW}}~). However, some positive residuals are
still found in the Fe line band, suggesting the presence of an
emission feature around 5~keV. We then added a narrow Gaussian (with
fixed zero width) and we detected emission features at $5.4$~keV in
obs.~1 and at $5.2$~keV in obs.~2 and 3. The narrow line EW is small
(less than 20~eV), but its inclusion in the model improves the
statistic ($\Delta\chi^2 \simeq 9-12$ depending on the observation
with two less degrees of freedom). Its possible origin is discussed
below. The results of the spectral analysis are reported in Table~1 as
{\small{MODEL~1}}. Taking into account the three observations, the
effective neutral hydrogen column density is found to be $N_H = (~0.5
\pm 0.1)\times 10^{22}$~atoms~cm$^{-2}$.

In Fig.~1, to demonstrate the presence of the broad Fe line in the
data, we show the data/model ratio obtained ignoring the 4-8~keV
band in fitting the model. The data have then be re--inserted to show
the broad residuals indicating that a relativistic Fe emission line is
present in the data, as detected with {\it{XMM--Newton}} by Miller et
al. (2002).

\subsubsection{The broad Fe line}

The {\small{LAOR}} model best--fit parameters indicate the presence of
a relatively strong and highly relativistic iron line with equivalent
width of about $EW \simeq 300$~eV. The
line energy in obs.~2 and 3 indicates a high level ionisation, while
in obs. 1 it suggests emission from less ionised iron. The emissivity
profile on the accretion disc is well described by a steep power law
with index $\beta \simeq 4-5$. This steep emissivity is consistent
with previous measurements with {\it{XMM--Newton}} \cite{miller02a}
and strongly suggests that the accretion disc is illuminated by a
centrally concentrated source of primary hard X--rays that strongly
irradiates the inner disc region.

The Fe emission line profile indicates emission from an accretion
disc extending down to about $2~r_g$. Such a small inner disc radius
is not consistent with a non--rotating Schwarzschild black hole if the
accretion disc is assumed to extend down to the marginal stable orbit
($6~r_g$ for a Schwarzschild black hole). On the other hand, since the
marginal stable orbit for a maximally rotating Kerr black hole has a
radius of about $1.24~r_g$, the data suggest the presence of a central
rapidly rotating Kerr black hole. However, the measure of an inner
disc radius smaller than $6~r_g$ does not necessary mean that the
black hole must be spinning. Indeed, Reynolds \& Begelman (1997)
showed that the line profile from an accretion disc around a
Schwarzschild black hole can be very similar to the one computed in
the Kerr spacetime, if emission within the marginal stable orbit is
considered (but see Young, Ross \& Fabian 1998). 

\subsubsection{The features around 5~keV}

As discussed above, a narrow emission line at $5.4\pm 0.1$~keV is
formally required in obs.~1, while the line energy is $5.2\pm 0.2$~keV
in obs.~2 and 3. Although the measured EW is low, these features worth
a brief comment. More than a single interpretation is of course
possible both as emission around 5.2--5.4~keV or as absorption at
slightly higher energies (for a similar feature in NGC~3516 see Nandra
et al. 1999). Narrow emission lines with $E < 6.4$~keV are also seen
in active galaxies (e.g. Turner et al.  2002, 2004; Guainazzi 2003;
Yaqoob et al. 2003; Dov\v{c}iak et al. 2004). The 5.2--5.4~keV
features in XTE~J1650--500 could part of the main broad
relativistic line. If, as it is likely, the emissivity on the disc is
more complex than a simple power law, it is possible that some small
features of the line profile are not well described by the
{\small{LAOR}} model. Redshifted iron emission from an outflow,
absorption due to resonant scattering in infalling material, or
spallation of iron producing a fluorescent Cr emission line (Skibo
1997) provide other interesting possibilities.

On the other hand, since the Fe line profile is very broad, an
explanation in terms of emission from the inner accreting flow seems
likely. It ispossible that the observed features are due to redshifted
Fe K$_\alpha$ emission line from the accretion disc, superimposed to
the main broad Fe line component. As an example, a strong flare
originating very close to the disc surface will illuminate only a very
small region of the disc. The emissivity profile would thus comprise
two components, a power law emissivity due to the illumination of the
whole disc by the corona emission (producing the broad Fe line) and a
small bump at the radius where the flare is active.  If this is the
case, we would mainly observe the blue horn of the emission line
associated with the flare, the red wing being faint and confused with
that of the main broad Fe line component. We tested this possibility
by replacing the narrow Gaussian with a second {\small{LAOR}} line
with energy and inclination tied to that of the main component and
found results comparable to those reported in Table~1 if the
5.2-5.4~keV line comes from a narrow ring around 2--3~$r_g$. Since the
quality of the data does not allow to distinguish between the
different possibilities, we will not discuss them any further. We
however include in our modelling a Gaussian emission line with fixed
zero width at 5.2-5.4~keV, depending on the observation.

\subsection{The broadband 1.5--200~keV spectrum of XTE~J1650--500}

In order to better constrain the Fe line parameters, a careful
analysis of the underlaying continuum is necessary. It is clear that a
power law and a {\small{SMEDGE}} model are simple, and probably
too rough, approximations to the hard spectral shape. In particular, the
presence of the Fe line strongly suggests that a reflection continuum
due to reprocessing of the primary X--rays by the accretion disc, is
present in the data. The 2.5--10~keV energy band is too limited to
constrain the reflection continuum, and for this reason we consider the
high energy data from the PDS instrument as well (in the range
13--200~keV). We also add some low energy data (1.5 to 3~keV) from the
LECS instrument, to better constrain the thermal component of the
spectrum. At lower energies, some absorption/emission features are
visible. We decide to ignore these features in this analysis that is
focused on the Fe complex and does not benefit much from the data
below 1.5~keV. 

The extension of the previous model ({\small{MODEL~1}}) to the
1.5--200~keV band does not provide a very good description to the
data. The main reason is the presence of spectral curvature and
positive residuals above 20~keV. The shape of the residuals suggests
the presence of a Compton hump due to a reflection component. Because
of the presence of these residuals and of the relativistic Fe emission
line, and anticipating the presence of a ionised disc as suggested by
the Fe line rest energies (see Table~1), we add a reflection component
({\small{PEXRIV}}, from Magdziarz \& Zdziarski 1995) to our model, and
remove the phenomenological {\small{SMEDGE}} model. 

Moreover, as all the components emitted from the accretion disc should
be affected by the same relativistic effects, the reflection
component, the thermal disc black body component and a narrow (i.e.
with fixed zero width) Gaussian emission line are relativistically
``blurred'' by convolving it with the core of the {\small{LAOR}}
model. The width of the Gaussian emission line is fixed to zero
because the Fe line profile is dictated by the effects of the
relativistic blurring; our description of the Fe emission line is
completely equivalent to the one provided by the {\small{LAOR}} model.
The incident continuum is described by a power law with high energy
cutoff; the photon index, cutoff energy and normalisation are tied to
those of the {\small{PEXRIV}} model. We also include a (non--blurred)
narrow emission line to model the features at $5.2-5.4$~keV, as
discussed in the 2.5--10~keV analysis. 

All abundances in the
reflection model were fixed to solar values while the relative
normalisation between incident power law and reflection (the relative
reflection $R$) was let free to vary during the fit. The outer disc
radius was fixed at 400~$r_g$ and the inclination at 45 degrees, as
before. We notice also that the same model was used by Miller et al.
2002 to describe the September 13 {\it{XMM--Newton}} data. Hereafter
we shall refer to this model as {\small{MODEL~2}}.

This model works well on the data of obs.~2 and 3 ($\chi^2 /dof =
533.5 / 434$ and $\chi^2 /dof = 434.8 / 416 $, respectively ) but a
reasonable fit can be obtained for obs.~1 only up to about 60~keV
($\chi^2 /dof = 402.5 / 324 $). In obs.~1 a sharp feature is detected
around $35$~keV in the PDS data and is most likely instrumental. If
the associated energy band is ignored in fitting the spectrum, the
statistics is much better ($\chi^2 /dof = 317.8/314$) and fit
parameters are unaffected.  However, the main problem for obs.~1 is
that, if higher energy data above $60$~keV are included,
{\small{MODEL~2}} fails to give an appropriate description
of the spectrum.  We shall report now our results with
{\small{MODEL~2}}, restricting the energy band to 1.5--60~keV for
obs.~1, and discuss later in some detail the broadband spectrum for
this first observation. The results of the fits with {\small{MODEL~2}}
are reported in Table~2.
\begin{figure}
\begin{center}
\psfig{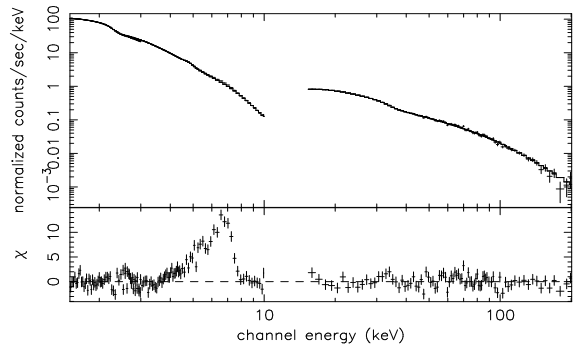}
\caption{The 1.5--200~keV spectrum of obs.~2 and the data/model ratio (in
  terms of $\sigma$) obtained with our {\small{MODEL~2}} when the
  4--8~keV band is ignored in the fitting. The data have
  been rebinned for visual clarity.}
\end{center}
\end{figure}

We measure an increase in the photon index from $\Gamma = 1.81$ in
obs.~1 to $\Gamma \simeq 2.1$ in obs.~2 and 3 which is consistent with the
general trend obtained by Rossi et al. (2003) from the analysis of
{\it{RXTE}} observations during the same outburst, and an inner black
body temperature of $0.6-0.7$~keV. The data support the presence of a
reflection component that contributes significantly to the hard flux,
especially during obs.~2 and 3 where the reflection fraction $R$ is
significantly greater than unity: the trend is of increasing $R$ with
time, from obs.~1 ($R \simeq 0.8$) to obs.~2 ($R\simeq 2.1$) and 3
($R\simeq 2.9$). The large values of the relative reflection measured
in obs.~2 and 3 indicate that the accretion disc is viewing more
illuminating radiation than we actually detect in the power law
component. The surface disc temperature and the ionisation parameter
($\xi$) of the reflection model vary in the range $1-9\times 10^6$~K
and $0.2-1.6\times 10^4$~erg~cm/s, respectively.  Although these
values are not well constrained by the data a trend of increasing
$\xi$ from obs.~1 to 3. is observed (see Table~2). Since the X--ray
flux decreases from obs~1 to 3 (see Section~4 and Table~3) this
indicates that the ionisation parameter is not correlated with
flux. A possible explanation for the behaviour of R and $\xi$ is given
in Section~4.

Once again, the most relevant results concern the broad iron emission
line. In all observations, the line emission is consistent with the
disc extending down to about 2 gravitational radii suggesting the
presence of a rapidly rotating Kerr black hole. Replacing the
{\small{LAOR}} kernel with the {\small{DISKLINE}} one which describes
a non--rotating Schwarzschild black hole, and fixing the inner disc
radius at $6~r_g$ results in a worse description of the spectra (the
smallest variation is $\Delta\chi^2 = 17$ with one more degree of
freedom in obs.~3, while the largest occurs in obs.~1 with
$\Delta\chi^2 = 70$).  We stress again that $r_{\rm{in}}$ is measured
not only through the iron line but is the result of the fit on all the
components that are thought to originate from the accretion disc that
are blurred with relativistic effects, so that the significance of the
measured $r_{\rm{in}}$ is, in our opinion, enhanced. The
{\small{PEXRIV}} model (lacking comptonisation) predicts too sharp an
edge at high levels of ionisation and could thus require extreme
relativistic blurring at high ionisation to smooth it out. However,
the measured $\beta$ and $r_{\rm{in}}$ do not show any trend toward
extreme values as the ionisation parameter increases from obs.~1 to 3
by nearly one order of magnitude, so that we are confident about the
robustness of our results. The emissivity index and the inner disc
radius are both consistent with being constant from obs.~1 to 3.

The line EW is about 200~eV, depending on the
observation and it is smaller than in the analysis of the
2.5--10~keV data (see Table.~1). This is mainly due to the non negligible
contribution of the ionised reflection continuum in the Fe line
region. The emissivity profile of the disc is quite steep ($\beta
\simeq 3.5-3.8$) suggesting that the iron line emission is concentrated in
the inner regions of the accretion disc. Summarising, our results
confirm the evidence for the presence of a Kerr black hole rotating
close to its maximum possible angular momentum ($a = 0.998$) in
XTE~J1650--500 as shown in the {\it{XMM--Newton}} observation by
Miller et al.  (2002), although other interpretations may still be
viable \cite{rb97}. As an example, the broadband 1.5--200~keV spectrum
of obs.~2 is shown in Fig.~2: the model is fitted ignoring the Fe line
energy band between 4 and 8~keV. This energy range is then
re--inserted to show the residuals in the Fe line region.

\subsubsection{The 1.5--200~keV spectrum of observation~1}
\begin{center}
\begin{figure}
\psfig{figure=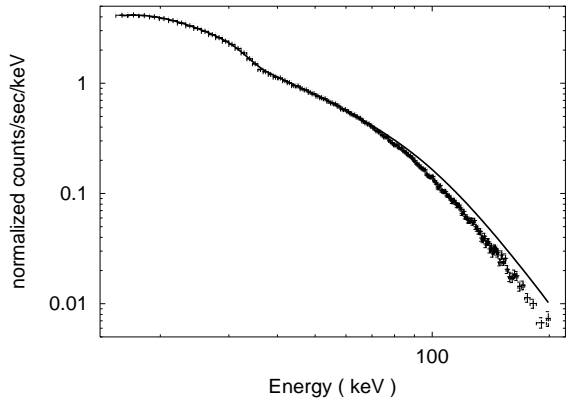,width=8.0cm,height=5.5cm,angle=-90}
\caption{The 15--200~keV spectrum of obs.~1 from the PDS instrument 
  demonstrates that our {\small{MODEL~2}} is inappropriate to describe
  the data above 60--70~keV. A much sharper cutoff than the one
  provided by the {\small{CUTOFFPL}} model is needed.}
\end{figure}
\end{center}

As already mentioned, {\small{MODEL~2}} does not provide a good
description of obs.~1 if data above 60~keV are included in the
analysis. The reason of the failure is clearly shown in Fig.~3, where
we show the PDS spectrum when the 1.5--60~keV best fit model for
obs.~1 is extended up to 200~keV. The data in obs.~1
clearly require a sharper cutoff above 60~keV than the one provided by
the exponentially decaying power law and reflection.  Even if the high
energy cutoff of the {\small{CUTOFFPL}} and {\small{PEXRIV}} models is
forced to assume low values, a reasonable fit can not be found. The
sharp cutoff is most likely the sign of a comptonised continuum which
is only roughly approximated by the exponentially cutoff power law.
For simplicity, we describe the sharp break with a simple
phenomenological model replacing the {\small{PEXRIV~+~CUTOFFPL}} model
with a {\small{BEXRIV~+~BROKENPL}}, i.e. assuming a broken power law
illuminating spectrum. The model provides a reasonable description to the
broadband spectrum of the first {\it{BeppoSAX}} observation (with
$\chi^2 /dof= 516.0 / 448$) without changing the results of the
1.5--60~keV analysis reported in Table~2. The difference is the
presence of a break energy at about 70~keV where the photon index
changes. Above that energy, the photon index steepens from
$\Gamma_{\rm{low}} \approx 1.8$ to $\Gamma_{\rm{high}} \approx 2.6$.
As before, ignoring the PDS sharp feature energy band around 35~keV
gives a much better fit with $\chi^2 / dof = 428.5 / 438$. The
spectral break is not visible in obs.~2 and 3 probably because of the lower 
signal to noise ratio in the PDS data that does not allow to constrain
the shape above 60~keV well enough.

\section{Fe line variability: evidence for light bending ?}

In this Section, we investigate the correlations between the
Power Law Component (PLC) that illuminates the accretion disc and
the reprocessed Reflection Dominated Component (RDC), i.e. reflection
continuum and the iron line.

In the three observations, we observe that the contribution of the
black body component to the total flux gradually increases with time,
while the PLC contribution shows the opposite trend. The 1.5--200~keV
flux of the black body component increases by about one order of
magnitude, while the PLC flux decreases by an even larger amount (see
Table~3). As already mentioned, 
the relative reflection $R$ increases by about a factor 3 
as the PLC decreases from obs.~1 to 3, suggesting that $R$ is
anti--correlated with the PLC flux. 
\begin{figure}
\rotatebox{270}{
\scalebox{0.3}{\includegraphics{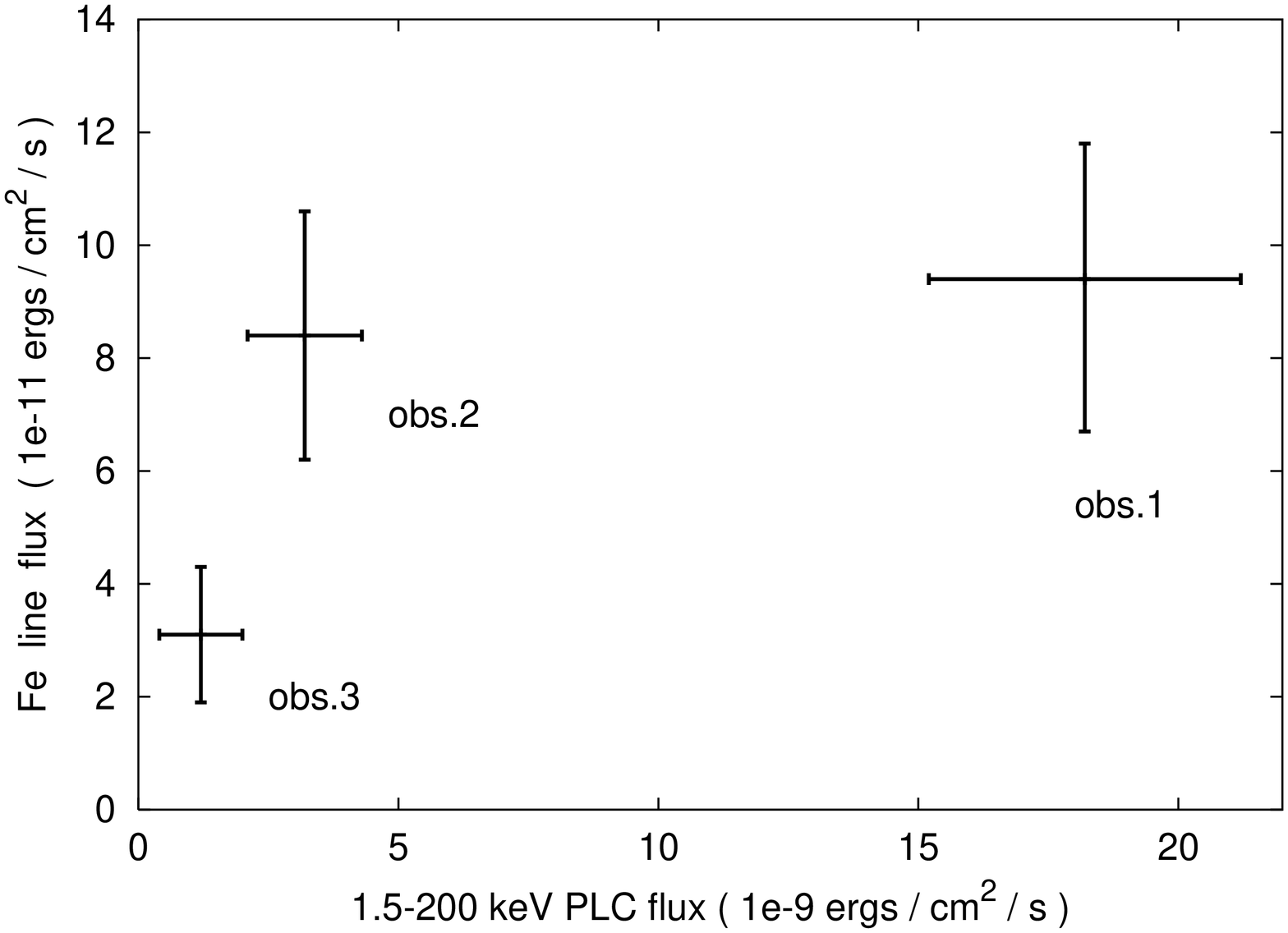}}}
\vspace{0.2cm}
\rotatebox{270}{
\scalebox{0.3}{\includegraphics{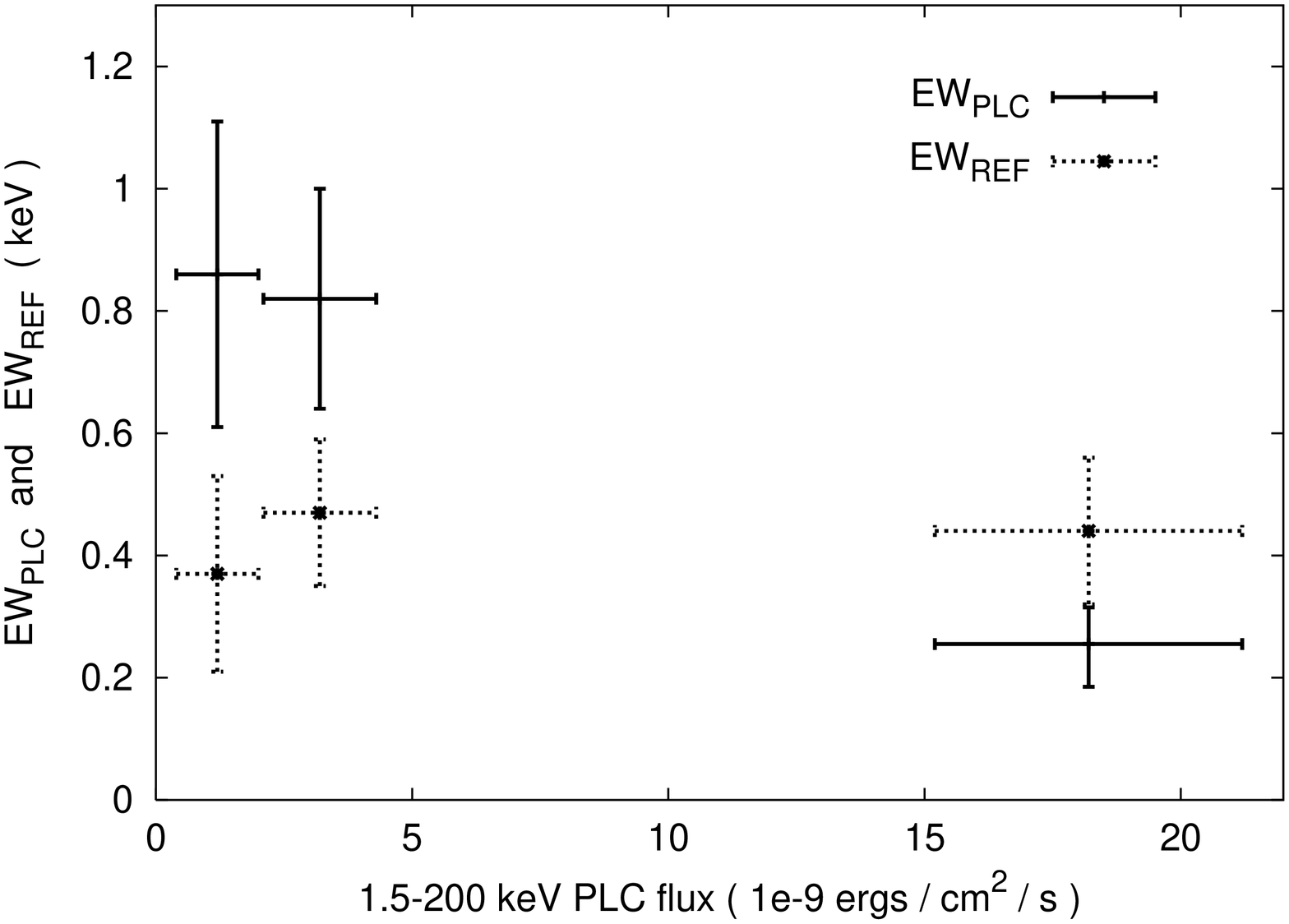}}}
\vspace{0.2cm}
\rotatebox{270}{
\scalebox{0.3}{\includegraphics{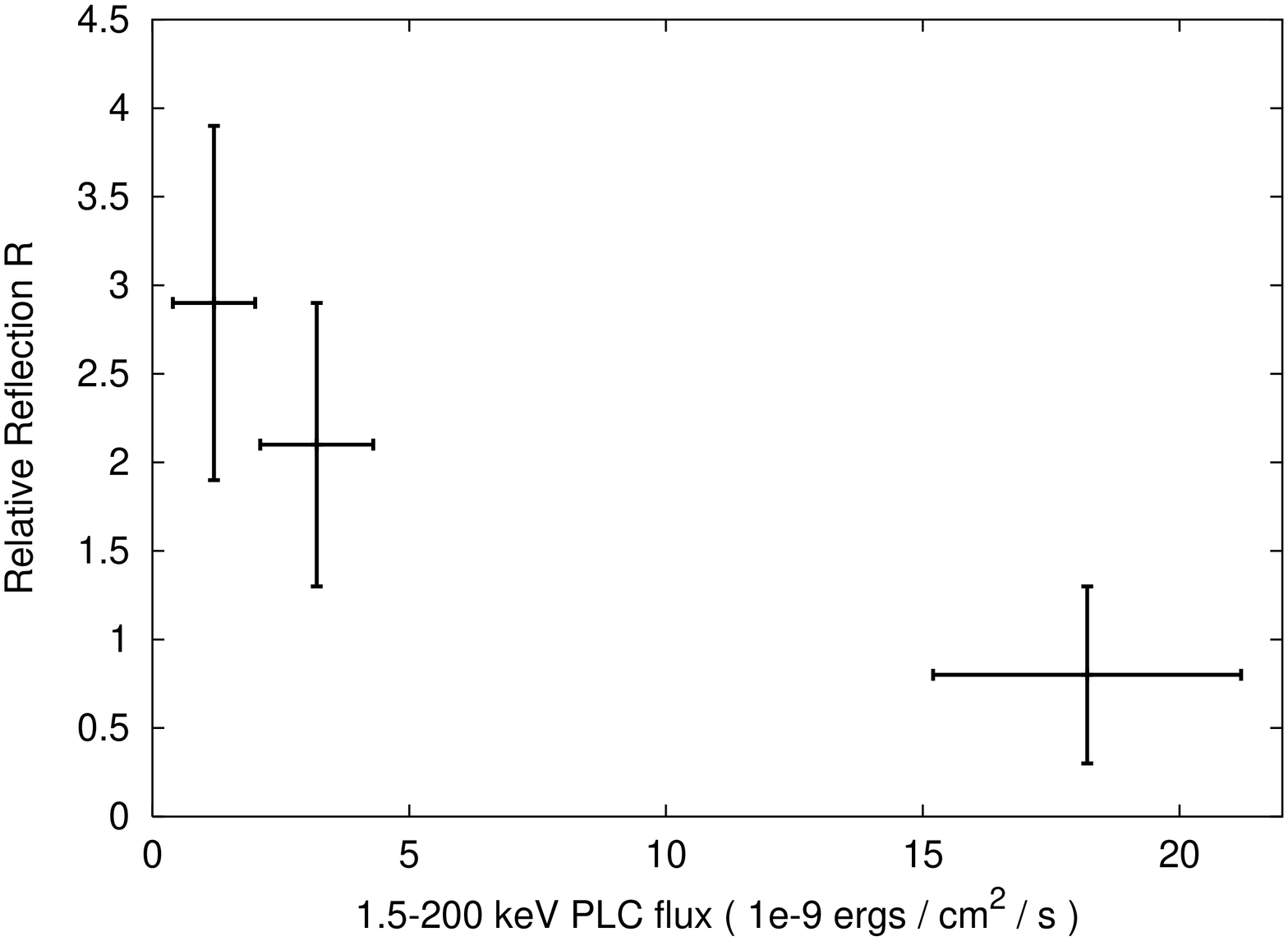}}}
\caption{ From top to bottom, we show, as a function of the
  1.5--200~keV flux of the PLC: i) the Fe line flux; ii) the Fe line
  EW computed with respect to the PLC only (EW$_{\rm{PLC}}$) and with 
  respect to the reflection continuum only (EW$_{\rm{REF}}$); iii) 
  the relative reflection R. The obs. number is showed in the top
  panel and refers to the other two
  figures as well. Results are discussed in the text.}
\label{corr}
\end{figure}

Our results are difficult to explain in the framework of the standard
view of reflection models because any RDC emitted from the accretion
disc is expected to respond to changes in the illuminating flux (the
PLC), so that $R$ should remain constant as the PLC flux changes.
However, this picture is probably oversimplified and fails (for
example) to account for the variability of the Seyfert 1 galaxy
MCG--6-30-15. As already pointed out by Miller et al. (2002), this AGN
exhibits some clear similarities with XTE~J1650--500.  In particular,
it is characterised by an extremely broad iron emission line (strongly
suggesting the presence of a Kerr black hole) and a steep emissivity
profile \cite{wetal01,fabetal02}.  MCG--6-30-15 exhibits also a
peculiar variability behaviour: in its normal states the spectral
variability can be accounted for by a phenomenological model
comprising a PLC that varies in normalisation and an almost constant
RDC (and iron line) \cite{sif02,fabv03,tum03}. In low flux states, the
iron line is correlated with the PLC \cite{rey03}.  This behaviour has
been successfully reproduced in terms of a light bending model in which
general relativistic effects strongly affect the emission in the near
vicinity of the central Kerr black hole \cite{mini03a,mf03b}.

The basic idea of the light bending model is that the PLC variability
is mostly due to changes in the height of a centrally concentrated
primary source of hard X--rays above the accretion disc rather than to
changes in its intrinsic luminosity. If the source height is small
(few gravitational radii), most of the radiation is bent onto the
accretion disc by the strong gravitational field of the central black
hole: this strongly reduces the observed PLC at infinity and enhances
the illumination of the disc. The Fe line (and any RDC) varies with
much smaller amplitude than the PLC and different regimes can be
identified in which the line is correlated with the PLC, almost
constant, or anti--correlated with the PLC. These regimes correspond
to low, medium and high PLC flux states.

A major prediction of the model is that the hard spectrum becomes more
and more reflection dominated as the PLC drops (i.e. as the height of
the primary source decreases) so that the relative reflection $R$
increases as the system goes into low PLC flux states. Large values of
the relative reflection are naturally expected in low PLC flux states.
Thus, if the PLC variability can be accounted for by light bending,
the increase of the relative reflection (and its large value) as the
PLC flux drops from obs.~1 to 3 can be understood. Moreover, as the
source height decreases and the PLC drops, the inner regions of disc
are more and more illuminated, so that a reasonable expectation is
that the ionisation parameter of the disc increases. This may
explain the trend of increasing ionisation parameter from obs.~1 to 3
(see Table~2).

The model predicts that the Fe line flux is correlated with the PLC in
low flux states, while it is almost constant in higher flux states.
Furthermore, the Fe line EW should correlate with the reflection
fraction, or in other words, the EW computed with respect to the
reflection continuum only (EW$_{\rm{REF}}$) is predicted be roughly
constant because, as in any reflection model, Fe line and RDC are
expected to vary together.  On the other hand, since the Fe line is
predicted to vary much less than the PLC, the EW computed with respect
to the PLC only (EW$_{\rm{PLC}}$) should be anti--correlated with the
PLC and saturate only at low fluxes where Fe line and power law
component are correlated.

As a test of the model, we then measure the Fe line flux, the PLC
flux, EW$_{\rm{REF}}$, and EW$_{\rm{PLC}}$. Results are shown in
Table~3 and Fig. \ref{corr}: as predicted by the light bending model,
the Fe line appears to be correlated with the PLC in low flux states
and to saturate at higher flux levels (top panel of Fig. \ref{corr}).
Moreover, EW$_{\rm{REF}}$ is almost constant, while EW$_{\rm{PLC}}$ is
anti--correlated with the PLC flux and possibly saturates at very low
PLC flux, in excellent agreement with the model (middle panel).  As
already mentioned, the relative reflection R is anti--correlated with
the PLC flux (bottom panel of Fig. \ref{corr}).

Clearly, three observations are not enough to establish correlations
properties with high level of significance.  However, the behaviour of
the iron line flux with respect to the PLC flux in XTE~J1650--500 has
been also measured with {\it{RXTE}} during the same 2001/2002 outburst
and results are reported by Rossi et al. (2003) where 176 pointed
observations are discussed.  The observed Fe~line~vs.~PLC behaviour is
remarkably similar to our present results and to the predictions of
the light bending model (see e.g. Figure 5 of Rossi et al. 2003 and
Figure 2 of Miniutti \& Fabian 2004). Of course, this does not mean
that light bending is really at work in this source and that other
models can not explain the correlation properties. However, the strong
similarity between theoretical predictions and observational data is
manifest.

\section{Discussion and Conclusions}

We present a broadband spectral analysis of the BHC XTE~J1650--500
during its 2001/2001 outburst. The three {\it{BeppoSAX}} observations
cover about 25 days starting from 2001 September 11, only a few days
after the peak of the outburst. We have observed a broad Fe line
profile in all three observations, confirming the detection with
{\it{XMM--Newton}} by Miller et al.  (2002a) during the very high
state on 2001 September 13. The Fe line profile we observe strongly
suggests that the accretion disc extends down to $r_{\rm{in}} \approx
2$, well within the limit of marginal stability for a non--rotating
Schwarzschild black hole, thus indicating that XTE~J1650--500 harbours
a rapidly rotating (possibly maximally rotating) Kerr black hole.

The Fe line emission is consistently associated with a strong
reflection spectral component from the accretion disc. The reflection
dominated spectrum is also distorted by the special and general
relativistic effects in the vicinity of the central black hole in the
same way as the Fe line. The emissivity profile of the disc seems to
be steeper than expected from standard disc models ($\beta \approx 3$)
suggesting that the primary source that illuminates the disc is
centrally concentrated.

From the first to the third observation, the power law component of
the spectrum drops by more than one order of magnitude. In the same
time, the relative reflection increases by a factor 3, suggesting that
the spectrum becomes more and more reflection dominated as the power
law flux decreases. We interpret this behaviour in terms of a light
bending model for the variability of X--ray sources that has been
presented elsewhere (see Miniutti et al 2003, 2004). The variability
of the Fe line flux and EW is also consistent with the predictions of
this model and with previous observations (Rossi et al. 2003).

\section*{Acknowledgements}
GM thanks Kazushi Iwasawa for constructive discussions and Alessandra
de Rosa for her help with {\it{BeppoSAX}} data analysis. We thank the
referee, Chris Done, for helpful comments and the
{\it{BeppoSAX}} Scientific Data Centre. GM thanks the PPARC for
support. ACF thanks the Royal Society for support. JMM thanks the NSF
for support through its AAPF program.

\clearpage


\begin{table*}
\centering
\caption{The 2.5--10~keV spectral analysis with {\small{MODEL~1}}. We
  do not report in the Table results on the narrow emission lines
  around $5$~keV (NL model in the Table). These are $E_{\rm{line}} = 5.4\pm 0.1$ in obs.~1, 
  and $E_{\rm{line}} = 5.2\pm 0.2$ in obs.~2 and 3, with an EW 
of $12\pm6$~eV in obs.~1 and 3 and an upper limit of $15$ in obs.~2.}
\begin{center}
\begin{tabular}{lccccccccc}                
\hline
\hline
{\small{MODEL~1}} & \multicolumn{9}{c}{{\small{PHABS~$\times$~SMEDGE~$\times$~[~PL~+~DISKBB~+~NL~+~LAOR~] }}} \\
\hline
 & $\Gamma$ & $KT_{\rm{in}}$ & $E_{\rm{line}}$ &$EW$
 & $\beta$ & $r_{\rm{in}}$ &  $E_{\rm{edge}}$ & $\tau_{\rm{max}}$ & $\chi^2 / dof$  \\
 \hline
\multicolumn{10}{l}{{ \small{2.5--10~keV SPECTRUM (~MECS~)}}}\\
\hline
obs.~1 & $1.77^{+0.05}_{-0.16}$ & $0.63 \pm 0.03$
& $6.60^{+0.11}_{-0.10} $ & $320 \pm 60 $ 
&$4.36^{+0.31}_{-0.19}$ & $1.88^{+0.19}_{-0.64}$
&$8.3^{+0.2}_{-0.3}$ & $0.5\pm 0.2$ &$146.1 / 147 $ \\ \\
obs.~2 & $2.69^{+0.38}_{-0.11}$ & $0.62 \pm 0.03$ &
$6.75^{+0.17}_{-0.20}$ & $340 \pm 80$ &
$4.28^{+0.34}_{-0.45}$ 
& $1.70^{+0.62}_{-0.46}$
& $8.3 \pm 0.4$ & $0.5\pm 0.2$ & $188.4 / 147$ \\ \\
obs.~3 & $2.6 \pm 0.3$ & $0.60 \pm 0.03$
&$6.87^{+0.1}_{-0.24}$ 
& $260 \pm 80$ & $5.2 \pm 0.3 $& $2.4^{+2.1}_{-1.0}$&
$8.2 \pm 0.4$& $0.6\pm 0.2$ & $138.2 / 147$\\ 
\hline
\hline
\end{tabular}
\end{center}
\end{table*}


\begin{table*}
\centering
\caption{The 1.5--200~keV spectral analysis with
  {\small{MODEL~2}}. The analysis has been restricted to the
  1.5--60~keV energy band for obs.~1 (see text for details).}
\begin{center}
\begin{tabular}{lccccccccc}                
\hline
\hline
{\small{MODEL~2}} & \multicolumn{9}{c}{{\small{PHABS~$\times$~[~CUTOFFPL$^1$~+NL1$^2$~+~RELBLUR$^3$~(~PEXRIV~+~DISKBB~+~NL2$^2$~)~] }}} \\
\hline
 & $\Gamma$ & $KT_{\rm{in}}$ & $E_{\rm{Fe}}$ &$EW^4$
 & $\beta$ & $r_{\rm{in}}$ &  $R$ & $\xi~(~\times 10^4~)$ & $\chi^2 / dof$  \\
\hline
\multicolumn{10}{l}{{ \small{1.5--200~keV SPECTRUM (~LECS~+~MECS~+~PDS~)}}}\\
\hline
obs.~1  & $1.81^{+0.08}_{-0.09}$ & $0.64^{+0.08}_{-0.05}$ &
$6.48^{+0.12}_{-0.06}$ & $160 \pm 60$ &
$3.51 \pm 0.15$ 
& $1.34^{+0.82}_{-0.10}$
& $0.80 \pm 0.5$ & $0.2 \pm 0.2$ & $402.5 /324$\\ (~1.5--60~keV~) &&&&&&&&&\\

obs. 2 & $2.13^{+0.03}_{-0.02}$ & $0.67 \pm 0.02$ &
$6.71^{+0.13}_{-0.09}$ & $250 \pm 80$ &
$3.8 \pm 0.2$ 
& $1.67^{+0.19}_{-0.21}$
& $2.1 \pm 0.5$ & $1.0^{+0.9}_{-0.2} $ & $533.5 /434$\\ \\

obs. 3 & $2.08 \pm 0.05$ & $0.63 \pm 0.04$
&$6.69^{+0.19}_{-0.16}$ 
& $190 \pm 90$ & $3.80^{+0.31}_{-0.22}$&$2.10^{+1.20}_{-0.70}$&
$2.9 \pm 1.0$& $1.6^{+1.2}_{-0.9} $ & $434.8/ 416$\\ 
\hline
\hline
\end{tabular}
\end{center}
\raggedright
$^1$ The parameters of the {\small{CUTOFFPL}} 
are tied to those of the
reflection model {\small{PEXRIV}} 
\\
\raggedright
$^{2}$ {\small{NL1}} is the narrow line around $5$~keV, while
{\small{NL2}} is affected by relativistic blurring and represents the
broad Fe line. In both cases, the width is fixed to
zero. Only results on the broad line are reported, those on the narrow
lines being identical to the previous 2.5--10~keV analysis (see
caption of Table~1). \\
\raggedright
$^3$ The relativistic blurring {\small{RELBLUR}} is provided by the
kernel of the {\small{LAOR}} model\\
\raggedright
$^{4}$ The Fe line EW (given in eV) is computed with respect to the total
underlaying continuum (in Table~3, different ways of computing it are
examined) 
\end{table*}


\begin{table*}
\centering
\caption{1.5--200~keV unabsorbed flux of the main spectral components are shown
  together with the relative reflection and the Fe line EW computed
  with respect to the PLC only and to the reflection continuum only. 
Fluxes are in units of $10^{-9}$~~erg/cm$^2$/s with the exception
of the Fe line flux which is in units of $10^{-11}$~~erg/cm$^2$/s. The
line EW is given in eV. }
\begin{center}
\begin{tabular}{lcccccccc}                
\hline
\hline
\multicolumn{8}{c}{{\small{1.5--200~keV FLUXES FROM THE BROADBAND BEST
    FIT MODELS}}} \\
\hline
 & $F_{\rm{tot}}$ & $F_{\rm{DISKBB}}$ & $F_{\rm{PLC}}$ &
 $F_{\rm{REF}}$ & $F_{\rm{line}}$ & $R$ & $EW_{\rm{PLC}}$ & $EW_{\rm{REF}}$ \\
\hline
obs. 1 & $25.0 \pm 7.0$ & $1.3 \pm 0.6 $  & $18.2 \pm 3.0$  
& $5.4\pm 3.2 $ & $9.4^{+2.4}_{-2.7}$ & $0.8 \pm 0.5$ &
$255^{+60}_{-70}$ & $440 \pm 120$\\ \\

obs. 2 & $16 \pm 4.0$ & $9.2 \pm 0.5$  & $3.2 \pm 1.1$  
&$3.6 \pm 1.2 $ & $8.4 \pm 2.2$ & $2.1 \pm 0.8 $  &
$820\pm 180$ & $470\pm 120$\\ \\

obs. 3 & $14.0 \pm 2.5$ & $11.8 \pm 0.5$   &$1.2 \pm 0.8$ & $1.8 \pm1.0$ 
& $3.1 \pm 1.2$ & $2.9 \pm 1.0$ 
& $860 \pm 250$ & $370 \pm 160$\\ 
\hline
\hline
\end{tabular}
\end{center}
\end{table*}


\label{lastpage}


\begin{thebibliography}{}

\bibitem[\protect\citename{Arnaud }1996]{arnaud}
Arnaud K.A., 1996, in ASP Conf. Ser. 101: Astronomical Data Analysis
Software and Systems V, 5, 17


\bibitem[\protect\citename{Boella, Butler \& Perola }1997a]{betal97a}
Boella G., Butler R.C., Perola G.C., 1997a, A\&AS, 112, 299

\bibitem[\protect\citename{Boella et al. }1997b]{betal97b}
Boella G. et al. 1997b, A\&AS, 112, 327

\bibitem[\protect\citename{Castro--Tirado et al. }2001]{ctetal01}
Castro--Tirado A.J., Kilmartin P., Gilmore A., Petterson O., Bond I.,
Yock P., Sanchez--Fernandez C., 2001, IAU Circ., 7707, 3

\bibitem[\protect\citename{Dov\v{c}iak et al. }2004]{dov04}
Dov\v{c}iak M., Bianchi S., Guanazzi M., Karas V., Matt G., 2004,
MNRAS in press, preprint (astro-ph/0401607)

\bibitem[\protect\citename{Fabian et al. }2002]{fabetal02}
Fabian A.C. et al., 2002, MNRAS, 335, L1

\bibitem[\protect\citename{Fabian \& Vaughan }2003]{fabv03}
Fabian A.C., Vaughan S., 2003, MNRAS, 340, L28

\bibitem[\protect\citename{Fiore, Guainazzi \& Grandi }1999]{fgg99}
  Fiore F., Guainazzi G., Grandi P., 1999, Cookbook for {\it{BeppoSAX}} NFI
  Spectral Analysis. SDC report
  (http://asdc.asi.it/bepposax/software/index.html)

\bibitem[\protect\citename{Frontera et al. }1997]{fetal97}
Frontera F., Costa E., dal Fiume F., Feroci M., Nicastro L., Orlandini
M., Palazzi E., Zavattini G., 1997, A\&AS, 112, 357

\bibitem[\protect\citename{Groot  et al. }1997]{groot01}
Groot P., Tingay S., Udalski A., Miller J., 2001, IAU Circ., 7708, 4

\bibitem[\protect\citename{Guainazzi  }2003]{gua03}
Guainazzi M., 2003, A\&A, 401, 903

\bibitem[\protect\citename{Homan et al. }2001]{homan01}
Homan J., Wijnands R., van der Klis M., Belloni T., van Paradijs J.,
Klein--Wolt M., Fender R., M\'endez M., 2001, ApJS, 132, 377

\bibitem[\protect\citename{Homan et al. }2003]{homan03}
Homan J., Klein--Wolt M., Rossi S., Miller J.M., Wijnands R., Belloni
T., van der Klis M., Lewin W.H.G., 2003, ApJ, 586, 1262

\bibitem[\protect\citename{Markwardt, Swank \& Smith }2001]{mss01}
Markwardt C., Swank J., Smith E., 2001, IAU Circ., 7707, 2

\bibitem[\protect\citename{Martocchia, Matt \& Karas }2002]{mmk02}
Martocchia A., Matt G., Karas V., 2002, A\&A 383, L23

\bibitem[\protect\citename{McClintock \& Remillard }2003]{mcrem03}
McClintock J.E., Remillard R.A., to appear in Compact Stellar X--ray
sources eds. W.H.G. Lewin and M. van der Klis, preprint (astro-ph/0306213)

\bibitem[\protect\citename{M\'endez \& van der Klis }1997]{mk97}
M\'endez M., van der Klis M., 1997, ApJ, 479, 926

\bibitem[\protect\citename{Miller et al. }2002a]{miller02a}
Miller J.M. et al., 2002a, ApJ, 570, L69

\bibitem[\protect\citename{Miller et al. }2002b]{miller02b}
Miller J.M., Fabian A.C., in'tZand J.J.M., Reynolds C.S., Wijnands R.,
Nowak M.A., Lewin W.H.G., 2002b, ApJ, 577, L15 

\bibitem[\protect\citename{Miller et al. }2002c]{miller02c}
Miller J.M. et al., 2002c, ApJ, 578, 348


\bibitem[\protect\citename{Miniutti et al. }2003]{mini03a}
Miniutti G., Fabian A.C., Goyder R., Lasenby A.N., 2003, MNRAS, 344, L22

\bibitem[\protect\citename{Miniutti \& Fabian }2004]{mf03b}
Miniutti G., Fabian A.C, 2004, MNRAS in press, preprint (astro-ph/0309064)

\bibitem[\protect\citename{Nandra et al. }1999]{nandra99}
Nandra K., George I.M., Mushotzky R.F., Turner T.J., Yaqoob T., 1999,
ApJ, 523, L17

\bibitem[\protect\citename{Parmar et al. }1997]{petal97}
Parmar A.N. et al. 1997, A\&AS, 122, 309

\bibitem[\protect\citename{Remillard }2001]{remi01}
Remillard R., 2001, IAU Circ., 7707, 1

\bibitem[\protect\citename{Revnivtsev \& Sunyaev }2001]{rs01}
Revnivtsev M., Sunyaev R., 2001, IAU Circ., 7715, 1

\bibitem[\protect\citename{Reynolds \& Begelman }1997]{rb97}
Reynolds C.S., Begelman M.C., 1997, ApJ, 488, 109

\bibitem[\protect\citename{Reynolds et al. }2003]{rey03}
Reynolds C.S., Wilms J., Begelman M.C., Staubert R., Kendziorra E., 
2004, MNRAS in press, preprint (astro-ph/0401305)

\bibitem[\protect\citename{Rossi et al. }2003]{rossietal03}
Rossi S., Homan J., Miller J.M., Belloni T., 2003, to appear in Proc. of the II
{\it{BeppoSAX}} Meeting, May 5--8, van den Heuvel E.P.J., in't Zand
J.J.M. and Wijers R.A.M.J. eds, Amsterdam, preprint (astro-ph/0309129)

\bibitem[\protect\citename{Shih, Iwasawa \& Fabian }2002]{sif02}
Shih D.C., Iwasawa K., Fabian A.C., 2002, MNRAS, 333, 687

\bibitem[\protect\citename{Skibo }1997]{sk97}
Skibo J.G., 1997, ApJ, 478, 522

\bibitem[\protect\citename{Tanaka et al. }1995]{tetal95}
Tanaka Y. et al., 1995, Nature, 375, 659

\bibitem[\protect\citename{Tanaka \& Lewin }1995]{tl95}
Tanaka Y., Lewin W., 1995, in ``X--ray Binaries'', Lewin W., van
Paradijs J., van den Heuvel E. eds., Cambridge University Press,
p. 126

\bibitem[\protect\citename{Taylor, Uttley \& McHardy }2003]{tum03}
Taylor R.D., Uttley P., McHardy I.M., 2003, MNRAS, 342, L31

\bibitem[\protect\citename{Turner et al. }2002]{tur02}
Turner T.J. et al., 2002, ApJ, 574, L22  

\bibitem[\protect\citename{Turner, Kraemer \& Reeves }2004]{tur04}
  Turner T.J., Kraemer S.B., Reeves J.N., 2004, ApJ in press, preprint
  (astro-ph/0310885)

\bibitem[\protect\citename{van der Klis }1995]{vdk95}
van der Klis M., 1995, in ``X--ray Binaries'', Lewin W., van
Paradijs J., van den Heuvel E. eds., Cambridge University Press,
p. 252

\bibitem[\protect\citename{Wijnands, Miller \& Lewin }2001]{wml01}
Wijnands R., Miller J.M., Lewin W.H., 2001, IAU Circ., 7715, 2

\bibitem[\protect\citename{Wilms et al. }2001]{wetal01}
Wilms J., Reynolds C.S., Begelman M.C., Reeves J., Molendi S.,
Staubert R., Kendziorra E., 2001, MNRAS, 328, L27

\bibitem[\protect\citename{Yaqoob et al. }2003]{yaq03}
Yaqoob T., George I.M., Kallman T.R., Padmanabhan U., Weaver K.A.,
Turner T.J., 2003, ApJ, 596, 85

\bibitem[\protect\citename{Young, Ross \& Fabian }1998]{yrf98}
Young A.J., Ross R.R., Fabian A.C., 1998, MNRAS, 300, L11

\end{thebibliography}
\end{document}